\documentclass[elsarticle,preprint,superscriptaddress,10pt]{revtex4}
\usepackage{amsfonts}
\usepackage{amsmath}
\usepackage{amssymb}
\usepackage{graphicx}
\usepackage{helvet}
\usepackage{hyperref}
\bibliography{aipsamp}
\begin{document}
\title{Synchronization of spatio-temporal semiconductor lasers and its application in
color image encryption}
\author{S. Banerjee}
\email{santo.banerjee@polito.it}
\affiliation{Dipartimento di Matematica, Politecnico di Torino, Corso Duca degli
Abruzzi 24, 10129 Torino, Italy}
\affiliation{Micro and Nanotechnology Unit, Techfab s.r.l., Chivasso, Italy}
\author{L. Rondoni}
\email{lamberto.rondoni@polito.it}
\affiliation{Dipartimento di Matematica and INFN, Politecnico di Torino, Corso Duca degli
Abruzzi 24, 10129 Torino, Italy}
\author{S. Mukhopadhyay}
\affiliation{Army Institute of Management, Judge's Court Road, Alipore, Kolkata-700 027, India}
\author{A. P. Misra}
\email{apmisra@visva-bharati.ac.in}
\affiliation{Department of Physics, Ume{\aa } University, SE--901 87 Ume{\aa }, Sweden.}
\affiliation{Department of Mathematics, Siksha Bhavana, Visva-Bharati University,
Santiniketan-731 235, India}
\received{13 October 2010}
\revised{22 November 2010}
\accepted{22 December 2010}
\begin{abstract}
Optical chaos is a topic of current research characterized by high-dimensional
nonlinearity which is attributed to the delay-induced dynamics, high
bandwidth and easy modular implementation of optical feedback.
In light of these facts, which adds enough {\it confusion} and {\it diffusion} properties for secure
communications, we explore the synchronization phenomena in spatiotemporal
semiconductor laser systems. The novel system is used in a two-phase
colored image encryption process. The high-dimensional chaotic attractor generated by
the system produces a completely randomized chaotic time series, which is
ideal in the secure encoding of messages. The scheme thus illustrated is
a two-phase encryption method, which provides sufficiently high confusion and
diffusion properties of chaotic cryptosystem employed with unique
data sets of processed chaotic sequences. In this novel method of cryptography, the chaotic phase masks are represented as images using the chaotic sequences as the elements of the image. The scheme
drastically permutes the positions of the picture elements. The next
additional layer of security further alters the statistical information of the
original image to a great extent along the three-color planes. The
intermediate results during encryption demonstrate the infeasibility for an
unauthorized user to decipher the cipher image. Exhaustive statistical tests conducted validate that the scheme is robust against noise and resistant to common attacks due to the double shield of encryption and the infinite dimensionality
of the relevant system of partial differential equations.
\end{abstract}
\keywords{Semi conductor laser, spatiotemporal chaos, synchronization, image cryptography, chaotic sequences, chaotic phase mask}
\maketitle


\section{Introduction}

The immense proliferation of internet and the advancement in
telecommunication has led to vulnerability problems in securely communicating the information. Confidentiality of data and information are thus an area of
concern for many researchers. An efficient and faster new breed of
encryption technique operating in the physical layer of the transmission
system is lately creating a lot of buzz. Quantum Cryptography and Chaos
Cryptography are subtle deviations from the traditional methods of encryption techniques
which are lately among the top research flurry. The first primarily focuses on quantum
key distribution. The second method deals with the scrambling and transfer of
encoded messages at a very high encryption data rate up to several tens of
Gb/s, unlike other software based traditional schemes. This method of
chaos cryptography is the outcome of the phenomenal work of chaos
synchronization demonstrated by Pecora and Carroll [1]. This bifurcation from
the traditional methods of encryption { realized} with the synchronization of two coupled
chaotic trajectories opened a whole new dimension for cryptographic
applications [2-7]. Its principle of operation relies primarily on the fact that
the chaotic nonlinear oscillator plays the role of a broadband signal
generator and the chaotic waveform can mask the information. The most distinct
feature of chaotic dynamics is that they possess ergodicity [8], sensitivity
to initial condition, randomness and mixing. These are the key properties
which are exploited in communication theory for secure transmission of data.
The experiment conducted by Cuomo and Oppenheim [9] demonstrated the
feasibility of such a communication scheme suggesting that the
unpredictability of chaotic oscillations of synchronized systems can be used effectively
to encrypt information-bearing signals, while their deterministic nature can
be used in decryption. In communication theory, a waveform which is coupled
with another can be formed as a carrier over the communication channel. Here,
due to chaos synchronization, the receiver end can retrieve the embedded
information. 

On the other hand, in order to design an efficient {cryptographic} system, it is
imperative to select a system which is { of high dimension} and remains chaotic in a continuous range of the parameter space [10]. Moreover, the encryption methodology used to develop an
efficient cryptic system should incorporate properties like a huge key space
[11], which { should be} very sensitive to any kind of perturbations and immune from
statistical attacks. Chaos based cryptography renders the cipher
deterministically disordered having a strong dependence on even minimal
variations of initial conditions and parameter values, and makes them resistant to
most of the attacks.

In apropos to the above discussion, another secure method of communication can
be realized at the physical layer itself by using chaotic carriers obtained
from semiconductor lasers to encode an information. Reference [12] elicits
that these nonlinear systems on synchronization { have tremendous potential to} achieve
enhanced security { where} they play the role of emitter and receiver system { when subjected to} a high injection strength [13].
It is well known that, higher the complexity of the chaotic carrier the more difficult { will be} to decode
the message without the appropriate receiver because of the high frequencies and the large number of degrees of freedom of the chaotic carrier. It
is worth noting that { the} chaotic carriers in semiconductor diode lasers provide a
broad spectrum in which the message can be hidden. With the advent of high
speed telecommunications { and advancements in computer processing [14] technology}, the demand for faster communication is fulfilled
with these laser systems which are increasingly used in major applications.

Researchers are now looking at ways to exploit lasers with chaotically
fluctuating signals, to add an extra layer of privacy to messages sent over
fiber-optic lines. Such optical systems will generate a chaotic carrier with
considerable high dimensionality which will enhance the security level
together with very high transmission rates. These higher dimensional
nonlinear devices exhibit nonlinear dynamical behavior with fast irregular
pulsations of the optical power, or wavelength hopping, with bandwidth
ranging from a few giga-hertz to tens of giga-hertz, large correlation
dimension of chaotic carriers [15,16] preventing linear filtering and
frequency-domain analysis, complexity and unpredictability of the
chaotic carriers, compact and integrated devices, high dimension due to the
delay-induced dynamics [17] etc. In recent years, experiments were conducted
with optical chaos based communication involving fiber lasers [18] and
semiconductor lasers [19]. Then Argyris et al [20] in 2005 established high
speed optical communication of 1Gbps over a commercial fiber-optic channel. This was a
pivotal point in improved communication over other commercial and common used modes of data
transmissions. The chaotic optical setup generated an optical carrier to
encode and transmit a message over a communication distance of $120$ km
achieving a robust transmission system, immune to perturbations and channel
noise. A major requisite for optical cryptography is to convert the message to be sent into an optical signal.
\begin{figure}[ptb]
\includegraphics[width=6in, height=4.0in,trim=0.0in 1.5in 0in 2in]{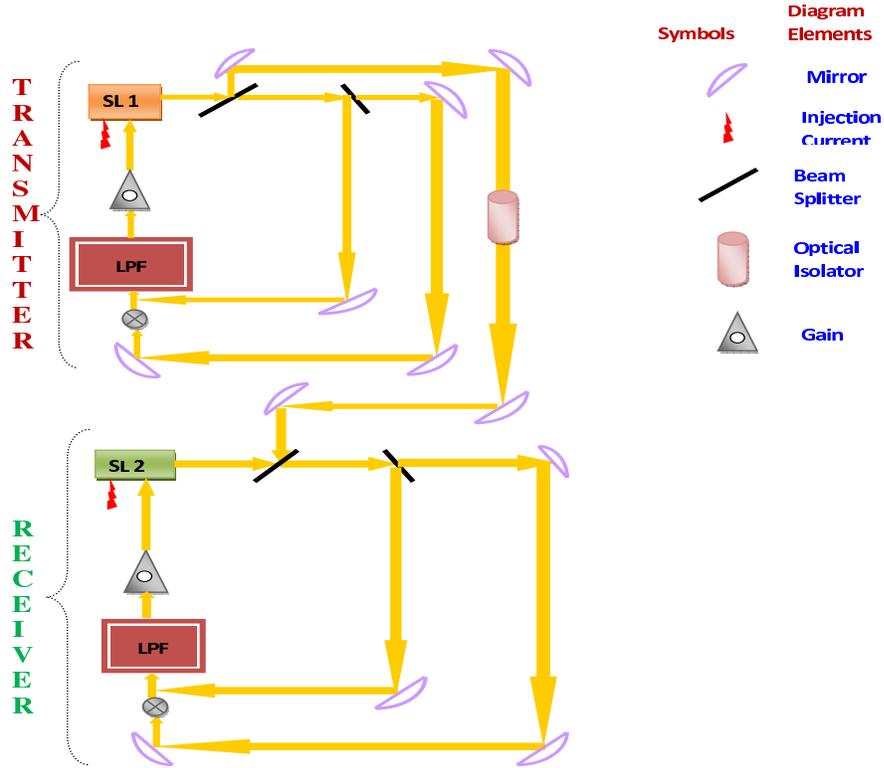}\caption{ (color online) Schematic diagram
depicting communication scheme with two semiconductor lasers subjected
to optical feedback  with the emitter and receiver lasers being coupled
unidirectionally. }%
\end{figure}

In order to realize secure communications using semiconductor lasers, the
emitter must exhibit a chaotic regime which is subjected to chaotic
oscillations by either injection from another source [21] or by reflection
from an external mirror [22]. This can be achieved by when the emitter is
either subjected to an optical or an electro-optical feedback. Our system
is used as an optical feedback, illustrated in \ Fig.1, with emitter and receiver semiconductor lasers unidirectionally coupled. The setup consists of semiconductor lasers which are
biased above a threshold level. These are then subjected to optical feedback
using external mirrors. The receiver operates as the emitter under similar
conditions in a closed loop or under CW when decoupled by open loop
scheme. A piezo transducer is employed when working in a closed loop scheme, so
as to match the external cavities. The optical signal thus produced is fed into
a laser that passes it along the laser beam. Then chaotic
synchronization occurs between the two spatially separated semiconductor lasers wherein the receiver side tunes its oscillations to that of the emitter
but suppressing the hidden information. At this stage, the message is
encrypted into the chaotic carrier. 

Synchronization basically implies that the receiver can reproduce the exact
chaotic irregular orbit of the emitter, i.e. the transmitter and the receiver
both generate the same chaotic signal by reproducing the irregular time
evolution of the same chaotic carrier, which traces a chaotic orbit at the
emitter end. Then the chaotic carrier wave within which the message is
embedded is transmitted from the emitter laser to an almost identical laser,
which forms the receiver section. The output of the receiving laser is tuned
to match the incoming signal from the transmitter. During
decryption, the laser in the receiver section only recognizes the chaotic
mask of the incoming information-bearing signal and extracts the originally transmitted message. It extracts the message by
subtracting the emitter laser beam (the input), which contains the message
from the identical carrier (output) generated at the receiver. 

Considerable research [23-25] shows how efficient optical feedback
laser systems are. In 1997, the authors [26] applied synchronization of the
optical chaos from external reflection technique based on optical feedback for
a robust cryptographic application by chaotic shift-keying, which is resistant
to external perturbations and noise. There have been other significant
successful attempts in chaos based optical communication and its application
in cryptography [27-34].  However, there are other issues to
be addressed regarding the reliability and efficiency of optical cryptography, concerning parameter mismatch, noise, dispersion and  modulation index of
chaotic carriers. Note, that the parameter mismatch has to be prevented since
the correct decoding at the receiver section intrinsically depends on the tuning
accuracy of the master and slave semiconductor lasers [35]. To
facilitate decryption the chaotic system possesses identical parameters and
operating conditions. If there is any mismatch, the devices will not be
synchronized and hence will not fulfill the goal of cryptography. Moreover,
the scheme has to overcome the hurdles imposed by the high-frequency together
with the high dimensionality of the chaotic signal.

Another important issue is the noise produced in the lasers or in the channel
communication. Such noise will hamper the accuracy of the synchronization due
to which the synchronization succumbs to destabilization. Then, the dispersion
factor of a chaotic laser is also an area of concern. The chaotic signal is
highly sensitive to dispersion, since it has a broad line width in the order of
Giga Hz in comparison to the output radiation of a solitary laser which has a
line width of only few Mega-Hz. Therefore, due to the increased line width,
the fiber communication links should be shorter than the traditional ones [36]
to facilitate propagation by chaotic semiconductor lasers. Lastly, the
modulation index of the encryption should be small to achieve enhanced security.

In this paper, we have {endeavored} to use synchronized optical feedback induced
chaotic dynamics for color image cryptography. The latter uses an external cavity
semiconductor laser on the transmitter side and a semiconductor laser with
optical injection on the receiver side. Image cryptography is achieved by
forming different masks derived from preprocessed chaotic sequences arranged
in the form of an image whose elements are chaotic
sequences.
\par { In the case of the two lasers, one should note that both systems are highly nonlinear and infinite dimensional in nature, even if one may only simulate a finite dimensional approximation, in a computer. The respective dynamics are also very sensitive to initial conditions, but it is possible to synchronize the two systems with a proper optical feedback mechanism. As we know, the synchronization of two continuous spatiotemporal PDE systems are quite a new subject of investigation. So, in our study, we have found interesting results which may motivate a new line of research in  cryptography.}

\par{ It is remarkable that, although both systems are infinite dimensional, they may easily synchronize with different initial and boundary conditions, using proper one way couplings. The experimental setup can be implemented as illustrated in the paper (Fig. 1) and the synchronization phenomenon between two lasers can then be used for practical purposes.}  {The cryptographic scheme is just an example of the possible applications in the field of communication, which profits from the features of the system. The differences between any synchronized chaos based cryptosystem
and our scheme are the following [1]: \footnotetext[1] {Detailed definitions of the quantities which we briefly mention here are given in the following sections.}}
\begin{enumerate}
  \item { The driving and the response systems are highly nonlinear and infinite dimensional in nature. Once the two systems synchronize, they remain synchronized forever. If we have synchronization at $t=t_{max}=5$ (dimensionless unit), they will remain the same for any $t_i > t_{max}$. Therefore, both the sender and receiver can start collecting data from any time after $t_{max}$. They can choose the data in an infinitely long time interval.}
  \item { We have three variables in our main PDE system [Eq.(1)] viz. the forward field $E_{+1}$, the backward field $E_{-1}$ and the normalized carrier density function $N$.  The keys used to encrypt/decrypt the image file are the variables $E_{\pm1}, E_{\pm2}$ and $N_1, N_2$ respectively. The interesting thing is that the driving signal from sender to receiver constitutes a one way coupling and contains only the electric fields (eq 4) of the driving system; it does not contain  any information about the carrier density. In this equation, $N_1$ as well as $N_2$ are not transmitted. Hence, the set of keys are}\\

{ Private/ secret key for transmitter $\rightarrow$ $N_1$\\
Private/ secret key for receiver $\rightarrow$ $N_2$\\
Private key for both parties $\rightarrow$ $t_{max}$\\
\\
Public key for the transmitter $\rightarrow$ $E_{+1},E_{-1}$\\
Public key for the receiver $\rightarrow$ $E_{+2}, E_{-2}$}\\
  \item { Therefore, any third party attempting to hack the signal sent from the transmitter to the receiver, will only get information about the electric field E but not  about the carrier density, which is used as our secret keys for encryption. They can try to reconstruct the phase space on the basis of the data obtained from the time series of the electric field E, but it is impossible to predict the nature of such a infinite dimensional PDE system.}
\end{enumerate}
{ These considerations motivated to exploit the potential of such a complex system in secure communication. The pre-requisites for any cryptographic application are met by our scheme: the system is high dimensional, guaranteeing an enormous key space, and is highly sensitive to minute perturbations, since it is spatio-temporally chaotic. This justifies consideration of our method for cryptographic applications. The system which is used to devise the algorithm for cryptographic application generates an infinite key space.}
\begin{itemize}
  \item { So after synchronization, and after the time $t_{max}$, a state $y_0$ is chosen, from which two sets consisting of infinitely many numbers, $Y^{1}$ and $Y^{2}$, are generated.}
  \item { A key should be large enough that a brute force attack (possible against any encryption algorithm) is infeasible, i.e. would take too long to execute. According to Shannon's work on information theory, in order to achieve absolute secrecy, it is necessary for the key length to be at least as large as the message to be transmitted and should be used once. This algorithm is called the One-time pad, also known as the Vignere cipher. Since longer symmetric keys require exponentially more work under brute force search, a sufficiently long symmetric key will prevent this type of attack. In our scheme, the key length is as large as the message which is transmitted [Eqs. (6),(7)]. Through the modulo operation, the length is restricted to the range $[0,M]$ or $[1,M+1]$. These processed keys are of the same length as that of the message and are used exactly once and therefore this organization is theoretically unbroken [4]. So, the advantage of the Vignere cipher is reaped by our scheme.Then in order to encode an image of length $M\times N$, we convert the set $Y^{1}$ and $Y^{2}$ into a chaotic sequence $K_1$ and $K_2$ respectively  containing the $M \times N$ elements as that of the plain image P of dimension $M\times N$. Each element of the set $K_1 and K_2$ are integers  in the range $[0,M]$ or $[1,M+1]$.  This conversion is through eq. 6 and eq 7. For example as illustrated later, with a $512 \times 512$ sized colored image $P$ we have restricted the key space. We obtain a set for P by transforming $y_0$ into an array of integers in the range $[0,512]$ from the processing steps in Eqs. (6) and (7).}

  \item { Although, chaos is an irregular motion, it is deterministic and therefore the original image can be completely recovered if the secret keys are exactly known. But in our scheme, two of the four secret keys, the carrier density and $t_{max}$ are not transmitted over the network and only the algorithm is assumed to be public. As a result, our scheme adheres to the Kerchkoff's principle for cryptographic algorithms.}

\end{itemize}

 The rest of the paper is organized as follows. In Section. II, we define a
set of spatiotemporal partial differential equations (PDEs) that describes the
dynamics of semiconductor lasers. The system exhibits chaos for some
particular values of the parameters. \ In Section. III, we have implemented the
synchronization phenomena in a set of coupled PDEs with linear feedback
coupling. The physical significance and the results are described in detail.
\ Section IV presents the encryption scheme based on the synchronized system of
PDEs. We study the security analysis of the proposed cryptosystem with some
statistical analysis in Section. V. Finally, Section. VI reports our conclusion.

\section{Spatiotemporal semiconductor lasers and their chaotic properties}

A spatiotemporal chaotic system is a spatially extended system that exhibits
spatiotemporal chaos (STC), i.e. higher disorder in both space and time. The disorder in space can be related by
rapid decay of spatial correlations whereas the temporal disorder can be measured by the
positive Lyapunov exponents.
Examples include the one dimensional coupled map lattices CML, which are discrete-time and discrete-space dynamical systems consisting of
nonlinear maps acting on the lattice sites, coupled with each other. Because of
the nonlinear dynamics of each local map and the diffusion due to
the spatial coupling, a CML exhibits STC [37]. Another
example of a system exhibiting spatiotemporal chaotic regime is the
semiconductor laser system. The advancements in telecommunications have opened
up a new arena in the study of nonlinear systems which include the high
nonlinearity demonstrated by laser systems. Individual chaotic systems are often
marred by the problem of rapid degradation of the chaotic dynamics in finite
computing precision [38]. It is in this area that spatiotemporal systems have
found an edge, for such chaotic systems have a sustained periodicity [36] with
a performance superiority in cryptography [39]. 

Investigations with optical systems have revealed that the dynamics of laser
systems with optical feedback depends on the size of the external cavity.
The semiconductor lasers have been found to oscillate with several
longitudinal modes simultaneously [40]. \ Such multimodal behavior increases
the dimensionality of the rate equations and has been extensively used in
applications, such as fiber couplers or compact discs. This is because
semiconductor lasers have short external cavities. To realize
this, one places an external reflector
which feeds back a
fraction of its delayed electromagnetic output into the active layer of the semiconductor laser. If the distance between the
reflector and the output facet of the semiconductor laser is $L_{ext}$, then
the round-trip time in the external cavity is $2L_{ext}/c$. A convenient model
of this configuration should be imperatively made in the field representation,
because the phase of the coherent laser radiation is strongly affected by the
delayed feedback. Therefore, the  system we consider takes the following form [41].%

\begin{align}
\pm\frac{\partial E_{\pm}}{\partial z}+\frac{\partial E_{\pm}}{\partial t} &
=i\left(  \frac{D_{p}}{L_{c}}\right)  \frac{\partial^{2}E_{\pm}}{\partial
z^{2}}-iWL_{c}E_{\pm}+\kappa\left[  \left(  N-1\right)  -i\alpha N\right]
E_{\pm},\nonumber\\
\frac{\partial N}{\partial t} &  =\left(  \frac{D_{f}T}{L_{c}^{2}}\right)
\frac{\partial^{2}N}{\partial z^{2}}+\frac{\Lambda T}{N_{0}}-\frac{T}{\tau
}N_{2}-aTE_{0}^{2}\left(  N-1\right)  \left(  \left\vert E_{+}\right\vert
^{2}+\left\vert E_{-}\right\vert ^{2}\right),\label{b1}%
\end{align}
where $E_{\pm}$ are the forward (backward) electric fields normalized by the
electric pump $E_{0}\sim10^{17}$ v/m and $N$ is the carrier density normalized
by $N_{0}$. The space and the time variables are normalized respectively by
$L_{0}\sim1\mu$ m and $T\sim0.01$ ns. The system of Eqs. (\ref{b1})
describes the weakly nonlinear dynamics of slowly varying two counter
propagating longitudinally high-frequency (hf) optical fields (traveling
waves) interacting with low-frequency (lf) density perturbations associated
with the electron-hole (total) charge carriers in the active semiconductor
lasers. In the case of a conservative system, the formation of optical
envelope solitons through the nonlinear interactions of such waves is most
important in the context of turbulence (in which energy is transferred to few
interacting wave modes) as well as in stable optical pulse propagation. Here
the amplitude of the optical wave envelope is modulated by the electrostatic
small but finite amplitude electron-hole density fluctuations in which $n_{l}$
represents the refractive index of the active layer, $D_{p}$ the diffraction
which provides higher-order dispersion. All other parameters and the
corresponding numerical values for exhibiting chaos are given in Table I.
Physically, since the coupling of the two laser strips, which is provided by
the overlapping evanescent optical fields and by diffusion of electron-holes,
can be changed by varying the inter-element distance, the dynamical behaviors
of the twin-strip lasers can show a transition from order to spatiotemporal
chaos of the optical field intensity [41]. Moreover, since the optical
properties are strongly affected by the local optical fields and the
properties of the charge carriers along the longitudinal directions, the
dynamical behaviors of them are to be determined in a self-consistent manner
as described above by the system of three PDEs.\begin{table}[th]
\caption{Laser Parameters for computations}%
\label{table:nonlin}
\centering
\begin{tabular}
[c]{lc}\hline\hline
Parameter & Value\#1\\[0.5ex]\hline
L (Cavity length) & 275 $\mu$m\\
w (Strip Width) & 5 $\mu$m\\
d (Active layer Thickness ) & 0.2 $\mu$m\\
$R_{1}$ (Reflexivity of the front mirror ) & 0.15\\
$R_{2}$ (Reflexivity of the back mirror ) & 0.95\\
$R_{3}$ (Reflexivity of the external cavity) & 0.95\\
$n_{l} $ (reflexive Index of the active layer ) & 3.5\\,trim=0.0in 0in 0in -1.5in
$\alpha$ (Linewidth enhancement gain) & 5\\
a (gain coefficient) & 1.5$\times10^{-16} cm^{2}$\\
$N_{0}$ (transparency concentrations) & 0.67$\times10^{18} cm^{3}$\\
$\lambda$ (Laser wave length) & 8000 nm\\
$\tau$ (nonradiative recombination time) & 5ns\\
$\tau_{ext}$ (Round trip time for the external cavity) & 830 ps\\
$L_{ext}$ (External cavity length) & 12 cm\\
$D_{f}$ (diffusion coefficient) & 30 $cm^{2}$/s\\
$D_{p}$ (Refraction Coefficient) & $10^{-5}$ m\\
$\eta$ (Injected efficiency) & 0.5\\
J (Injected current) & 80/95/110/130 mA\\
$\alpha_{int}$ (Internal loss) & \\
s (Surface recombination velocity) & $10^{6}$m/s\\
$\beta$ (output scaling factor) & 1\\
$\Gamma$ (Confinement factor) & 0.5\\
$n_{c}$ (refractive index of cladding) & 3\\
$n_{eff}$ (Effective index) & 3.5\\
$N_{0}$ (Career density at transparency) & $10^{24}m^{-3}$\\[1ex]\hline
\end{tabular}
\end{table} 
We numerically solve the Eqs. (\ref{b1}) by Runge-Kutta method with
initial conditions as [42,43]
\begin{equation}
E_{\pm}(z,0)=E_{\pm0}+\tilde{E}_{\pm1}\cos(kz)/\beta,\text{ }N=-\tilde{N}%
\cos(kz)/\beta,\label{i1}%
\end{equation}
where $k$ is the wave number of modulation, $\ \beta$ is a suitable constant
taken as $1/500$ to ensure that the perturbation is sufficiently small,
$E_{\pm0}=2$ and $\tilde{E}_{\pm1},$ $\tilde{N}$ are constants of the order of
unity. The numerical results are shown in Fig. 2 after the end of the
simulation at $t=100$. We increase the injected current from $J=80$ to $130$
mA with external cavity length $L_{ext}=15$ cm and round-trip time $\tau=830$
ps. We choose the injection efficiency $\eta=0.8$ to assume that $80\%$ of the
carrier reaches the active region. 
\begin{figure}[ptb]
\includegraphics[width=7.0in, height=4in,trim=0.0in 2in 0in 2in]{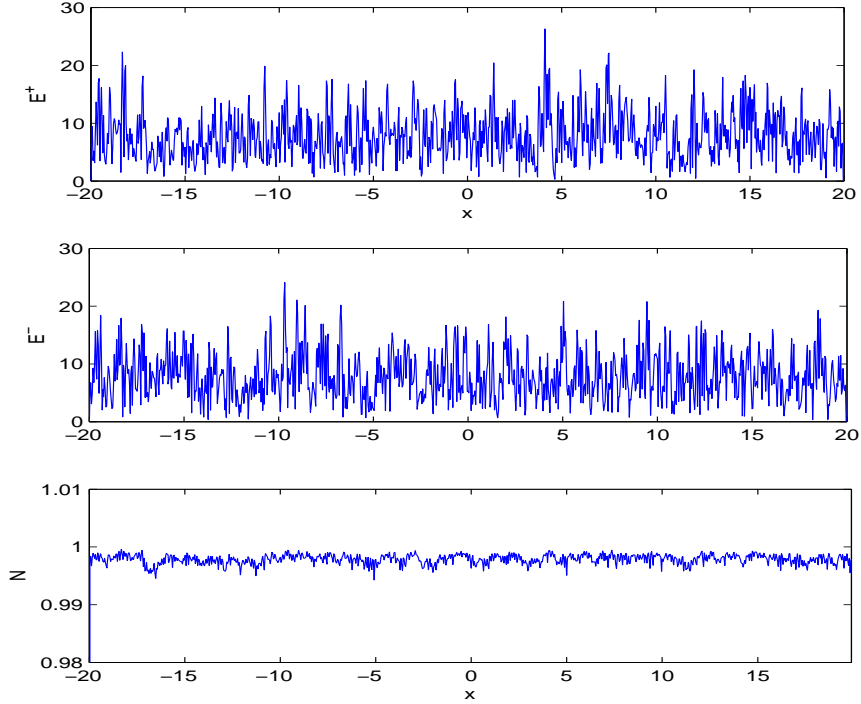}\caption{(color online) Numerical
solution of \ Eqs. (\ref{b1}) for forward and backward electric fields (see
the upper two panels) \ as well as the carrier density (lower panel)
exhibiting spatiotemporal chaos.}%
\end{figure}
From Fig. 2, we find that the forward as well as the backward
fields are chaotic in nature along with the density perturbation, and we can
optimize the maximum electric fields to obtain values greater than $|E_{\pm}|=20$ for both waves.

\section{Synchronization of spatiotemporal semiconductor lasers}

In this section we will synchronize the two nearly identical semiconductor
laser systems [\textit{c.f.} Eqs. (\ref{b1})] which exhibit chaos as
shown above. To this end, we use linear coupling terms with a coupling
coefficient $\epsilon$. The former { behaves} like a damping rate of the optical
fields in order to balance or manifest the nonlinear laser loss or gain, and
can be controlled through the variation of the coupling strength $\epsilon$.
Thus, we have the following system of equations where the suffix $1$ is used
to denote the driver, and $2$ the response system.%

\begin{align}
\pm\frac{\partial E_{\pm1}}{\partial z}+\frac{\partial E_{\pm1}}{\partial t}
&  =i\left(  \frac{D_{p}}{L_{c}}\right)  \frac{\partial^{2}E_{\pm1}}{\partial
z^{2}}-iWL_{c}E_{\pm1}+\kappa\left[  \left(  N-1\right)  -i\alpha N\right]
E_{\pm1},\nonumber\\
\frac{\partial N_{1}}{\partial t}  &  =\left(  \frac{D_{f}T}{L_{c}^{2}%
}\right)  \frac{\partial^{2}N_{1}}{\partial z^{2}}+\frac{\Lambda T}{N_{0}%
}-\frac{T}{\tau}N_{1}-aTE_{0}^{2}\left(  N_{1}-1\right)  \left(  \left\vert
E_{+1}\right\vert ^{2}+\left\vert E_{-1}\right\vert ^{2}\right)  , \label{s1}%
\end{align}

and
\begin{align}
\pm\frac{\partial E_{\pm2}}{\partial z}+\frac{\partial E_{\pm2}}{\partial t}
&  =i\left(  \frac{D_{p}}{L_{c}}\right)  \frac{\partial^{2}E_{\pm2}}{\partial
z^{2}}-iWL_{c}E_{\pm2}+\kappa\left[  \left(  N-1\right)  -i\alpha N\right]
E_{\pm2}+\epsilon\left(  E_{\pm1}-E_{\pm2}\right)  ,\nonumber\\
\frac{\partial N_{2}}{\partial t} &  =\left(  \frac{D_{f}T}{L_{c}^{2}}\right)
\frac{\partial^{2}N_{2}}{\partial z^{2}}+\frac{\Lambda T}{N_{0}}-\frac{T}%
{\tau}N_{2}-aTE_{0}^{2}\left(  N_{2}-1\right)  \left(  \left\vert
E_{+2}\right\vert ^{2}+\left\vert E_{-2}\right\vert ^{2}\right)  ,\label{s2}%
\end{align}
where $\kappa=\Gamma L_{c}N_{0}a,\epsilon=\zeta L_{c},L_{c}=Tc/n_{l}.$ Thus,
the two coupled multi-mode chaotic semiconductor lasers given by Eqs.
(\ref{s1}) and (\ref{s2}) describe the driver and response system respectively to
demonstrate chaotic communications in high-dimensional systems through
synchronization. This will enable the transmission of multiple messages
through a single or multiple scalar complex field. The synchronization of the
two systems (\ref{s1}) and (\ref{s2}), is achieved through the inclusion of
coupling terms $\propto\epsilon$ which, on the other way, \ describe
non-radiative laser loss or gain by the driver or response field due to power
reflectivity. \ It is, therefore, of importance to see whether synchronization
of such high-dimensional chaotic systems is robust to noise and/ or error
signals for transmitted messages. Using the similar initial conditions \ along
with $E_{\pm},N$ and their derivatives vanish at the boundaries, we integrate
the coupled systems (\ref{s1}) and (\ref{s2}) numerically. The simulation
results are displayed in Fig. 3. These basically represent the synchronization
error between the driving laser (\ref{s1}) and the response one (\ref{s2})
respectively at the most chaotic state. The initial values for numerical simulation are given in Eq. (2) with $E_{+10}=N_1=0.2$, $E_{+11}=1$, $E_{-10}=3$, $E_{-11}=1$.   We find that the corresponding errors
for the forward and backward electric fields are of the order $10^{-11}$ and
for the carrier density the difference is of the order $10^{-14}$. So, at the
synchronization state the system (\ref{s1}) can be considered as the
transmitter and the system (\ref{s2}) as the receiver one.

\begin{figure}[ptb]
\includegraphics[width=6in, height=4in,trim=0.0in 3in 0in 3.5in]{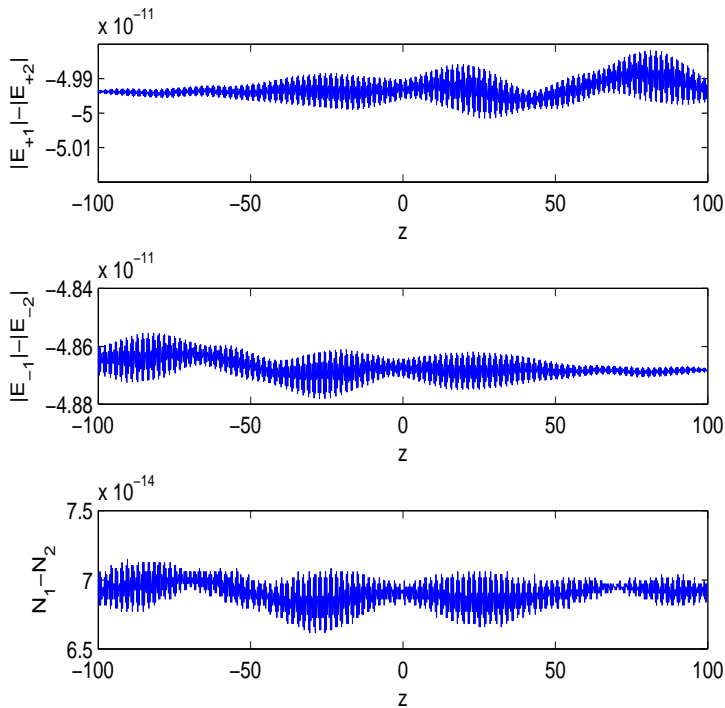}\caption{(color online)
Synchronization errors of coupled spatiotemporal semiconductor laser systems.
}%
\end{figure}

\section{Optical Chaos and application to image encryption}

This section focuses on the application of the synchronized space time laser systems to digital cryptography. Image encryption basically denotes the realignment of the pixels to different positions with a change  in the statistical value of the pixels. This permutation yields a disordered version of the image which is totally incoherent. In order to realize image cryptography, it is imperative to devise a transmitter and receiver section. It is in this arena that synchronization is utilized. For cryptographic encoding, the transmitter for our scheme uses
system (3) and the receiver is formed from system (4). Both of them are certain to chose values from the time series $E_{\pm1}$, $N_{1}$
and $E_{\pm2}$, $N_{2}$ respectively after synchronized space and time for the
proposed cryptographic scheme.The proposed method is simulated in Matlab 7.0 environment.

\subsection{Generation of chaos sequence}

A colored digital image $P(M,N,D)$ is basically composed of three sets of
two-dimensional (2D) matrices containing integers in the range of  $0-255$.
Each matrix contains $M$ number of rows and $N$ number of columns and has
three-color planes, $D$ namely Red (R), Green (G) and Blue (B) which are
concatenated along the third dimension $D$. Each picture element of the
three-dimensional (3D) matrix in $M\times N\times3$ format is actually a
triplet known as a {\it voxel}. It is analogous to its 2D monochrome image, the pixel. In this work, we would henceforth refer a picture
element for a colored image as a voxel. Due to the availability of fast
processing and high bandwidth technology, it is feasible to employ encryption
directly onto the 3D image instead of pre processing it to an indexed 2D monochrome image. This reduces considerable computational costs. Hence, the {\it plain image P}
containing $len=3MN$  voxels expressed by
\[
P=\{p_{1}=p_{1}^{R},p_{2}=p_{1}^{G},p_{3}=p_{1}^{B},p_{4}=p_{2}^{R}%
,p_{5}=p_{2}^{G},p_{6}=p_{2}^{B},\ldots,p_{i},\ldots,p_{len}=p_{MN}%
^{B}\}.\eqno {(5)}
\]

\subsection{The algorithm for image encryption}

In digital image cryptography, Confusion and diffusion are considered as
two distinct stages, both requiring image-scanning to obtain pixel/voxel values. In
our method, the original message undergoes a complete randomized permutation of the
positions in the diffusion phase and a drastic statistical change in the
confusion phase. The proposed method eliminates the necessity of duplicated
scanning thereby enhancing the encryption speed. At
first, the image is partitioned into blocks of voxels. Then STC is employed
to shuffle the blocks with a simultaneous change in the voxel values.
Also, an efficient method for generating pseudorandom numbers from STC is
suggested, which further increases the encryption speed. Theoretical analysis
and computer simulations confirm that the new algorithm is highly secure
and is very fast for practical image encryption.

\begin{figure}[ptb]
\includegraphics[width=6in, height=3.5in,trim=0.0in 1in 0in 2in]{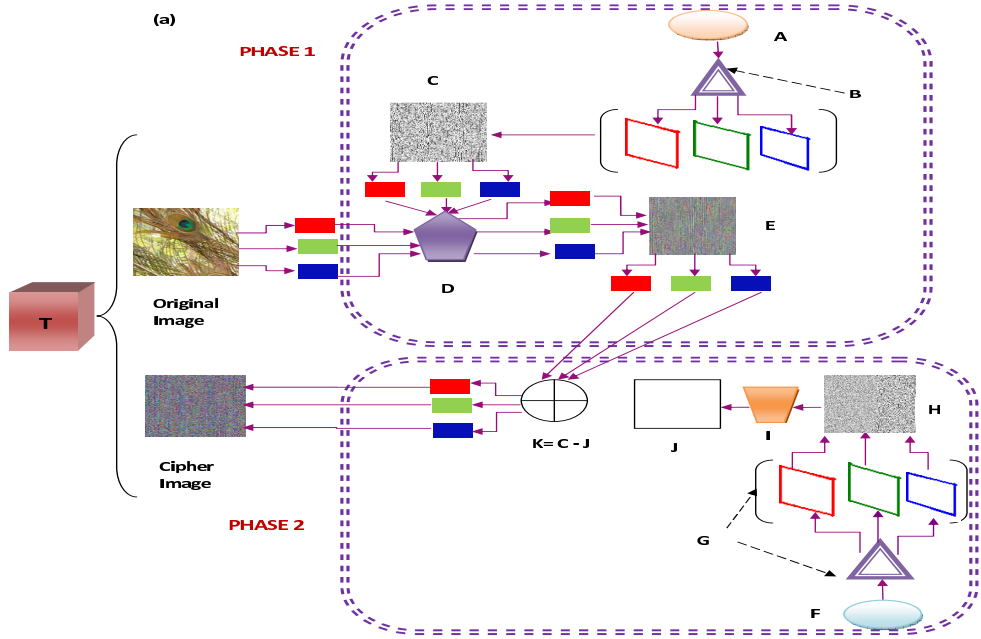}\caption{(color online) Illustration
of the encryption scheme: (a) Encryption by transmitter- T . Symbols :- A - Chaotic data set $Y^{1}$, \ B -
Transformation along the color planes to form $CSI^{1}$ , C - mask $CSI^{1}$,
D- Shuffling operation in encryption, E - The shuffled image T,  F - Chaotic
data set $Y_{2}$, G - Transformation along the color planes to form $CSI^{2}$,
H - the second encryption mask $CSI^{2}$, I - Transpose operation, J - The
transposed mask $CSI^{2^{\prime}}$, K - differential mask $CSI^{3}$.}%
\end{figure}
\begin{figure}[ptb]
\includegraphics[width=6in, height=4in,trim=0.0in 1in 0in 2in]{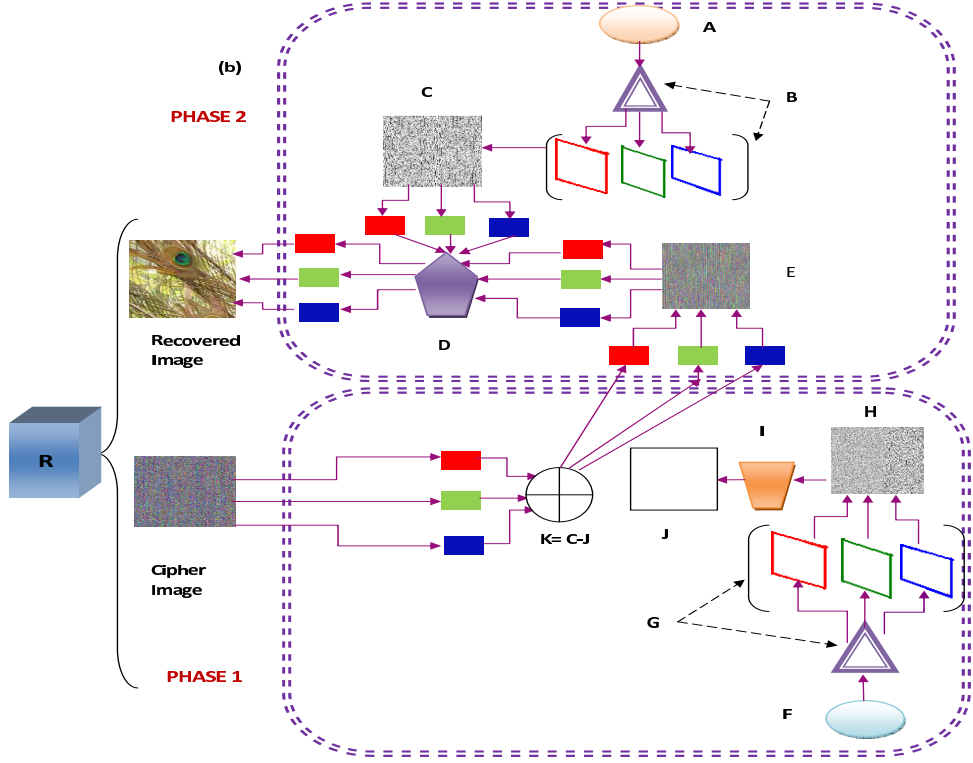}\caption{(color online) Illustration
of the decryption scheme: (b) Decryption in
Receiver section - R. Symbols :- A - Chaotic data set $Y^{1}$, B -
Transformation along the color planes to form $CSI^{1}$ , C - mask $CSI^{1}$,
\ D- Anti-shuffling in decryption, E - The shuffled image T, F - Chaotic data
set $Y_{2}$, G - Transformation along the color planes to form $CSI^{2}$, H -
the second encryption mask $CSI^{2}$, I - Transpose operation, J - The
transposed mask $CSI^{2^{\prime}}$, K - differential mask $CSI^{3}$. }%
\end{figure}
The stream cipher mode of encryption has been implemented at the bit level
with RGB image. We employ the advantages of ergodicity and sensitivity to
initial conditions of synchronized chaotic system which leads to the confusion
and diffusion properties required for secure cryptosystem. The entire cryptography scheme is illustrated in Figs. 4 and  5 and explained
as {\bf below}.\newline

\textit{Preprocessing Stages}:-\newline

\begin{enumerate}
\item :- The system under consideration is subjected to a maximum number of
iterations $t_{max}$.  An initial synchronized state $y_{0}$ is selected to
extract the chaotic sequences from variables $E_{\pm1}$, $N_{1}$ at the
transmitter end and $E_{\pm2}$, $N_{2}$ at the receiver section. Then, the i
-th chaotic data for permutation of the elements of the RGB image $P$ of size
$3\times M\times N$ is obtained from a synchronized data set $Y^{1}%
=\{y_{1}^{1},y_{2}^{1},\ldots,y_{\infty}^{1}\}$ and $Y^{2}=\{y_{1}^{2}%
,y_{2}^{2},\ldots,y_{\infty}^{2}\}$ respectively from $E_{\pm1}$, $N_{1}$ and
are preprocessed since the elements are real numbers. The scheme
employed is a duplex encryption technique. Hence, the two different data sets
of keys, namely $K^{1}=\{k_{1}^{1},k_{2}^{1},\ldots,k_{\infty}^{1}\}$ and
$K^{2}=\{k_{1}^{2},k_{2}^{2},\ldots,k_{\infty}^{2}\}$ that are used to encrypt the
image are generated through the simple algebraic relation\newline%
\[
k_{i}^{1}\leftarrow\text{integer}(\text{abs}(10^{3}\times|y_{i}^{1}%
|))\text{Mod}(M+1),\eqno {(6)}
\]
and%
\[
k_{i}^{2}\leftarrow\text{integer}(\text{abs}(10^{5}\times|y_{i}^{2}%
|))\text{Mod}(M+1).\eqno {(7)}
\]
Hence, $K^{1}$and $K^{2}\epsilon$ $[0,M]\newline$

\item :-

\begin{itemize}
\item Next we select a subset of integers from the processed data set $K^{1}$
and $K^{2}$from Eqs. (6) and (7) into $K^{1^{\prime}}$, $K^{2^{\prime}}$
respectively which will now be an array of $M\times N$ data elements with $M$
rows and $N$ columns and further process these through the following
rules\newline%
\[
K^{1^{\prime}}\leftarrow\text{reshape}(K^{1},M,N),\eqno {(8)}
\]%
\[
K^{2^{\prime}}\leftarrow\text{reshape}(K^{2},M,N).\eqno {(9)}
\]

where $\it{reshape}$ means to change the size of $K^{1}$ and $K^{2}$ to $M\times N$ array, from which the processed elements are extracted in a column wise fashion. The result of this operation will be an array containing an $M \times N$ number of elements.
\item $K^{1^{\prime}}$ and $K^{2^{\prime}}$ can be represented as an image. In our scheme, we have tiled the array which have resulted from Eqs (8) and Eqs (9) along the three color
planes by using the command {\it repmat(X,[P Q])} in Eqs. (10) and (11). This will produce a large array consisting of $P\times Q$ block of array where $X$ may be a multi dimensional array. Since, in this operation we generate an $1\times3$ tilings of copies of $K^{1^{\prime}}$ and $K^{1^{\prime}}$
respectively, we denote it as equivalent to an image matrix which we label
it as the Chaotic Sequence Image $CSI_{M,N,D}^{1}$, $CSI_{M,N,D}^{2}$ respectively. \newline%
\[
CSI_{M,N,D}^{1}\leftarrow\text{repmat}(K^{1^{\prime}}%
,[1\hskip2pt1\hskip2pt3]),\eqno {(10})
\]%
\[
CSI_{M,N,D}^{2}\leftarrow\text{repmat}(K^{2^{\prime}}%
,[1\hskip2pt1\hskip2pt3]).\eqno {(11})
\]
Here, $CSI^{1}$ will serve as the permutation mask [Fig. 6 (a)] in the diffusion
stage and $CSI^{2}$ {[Fig. 6(b)]} is used during the next phase of
encryption. Each element of these phase masks are composed of chaotic sequences and therefore are completely randomized themselves.\newline

\begin{figure}[ptb]
\includegraphics[width=6in, height=3in,trim=0.0in 4in 0in 4in]{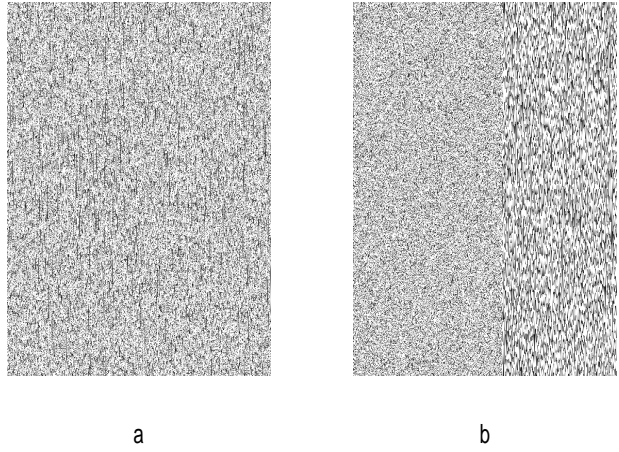}\caption{(color online) The masks
used in two phase cryptography: (a) Mask generated from first chaotic sequence
(b) Mask generated from second chaotic sequences. }%
\end{figure}
\end{itemize}

\item :- The $CSI^{2}$ is transposed into $CSI^{2^{\prime}}$
as\newline%
\[
CSI_{r^{^{\prime}},c^{^{\prime}},d^{^{\prime}}}^{2^{\prime}}\leftarrow CSI_{c,r,d}^{2},\eqno {(12)}
\]
where by $r^{^{\prime}}=\{1,2,3,\ldots,N\},$
$c^{^{\prime}}=\{1,2,3,\ldots,M\}$ and $d^{^{\prime}}=\{1,2,3\}$ Then the statistical difference between $CSI^{1}$ and $CSI^{2^{\prime}}$ is
computed as \newline%
\[
CSI^{3}\leftarrow\text{abs}(CSI^{1}-CSI^{2^{\prime}}%
,\hskip 2pt mod \hskip 2pt (M+1)).\eqno {(13)}
\]

\end{enumerate}

\textbf{Encryption:-}\newline\newline\textit{Position Shuffling :- Diffusion
phase}\newline\newline Step A:-\newline\newline Sort the elements of
$CSI_{r,c,d}^{1}$ in ascending order in accordance to their voxel values as a
vector where $r=\{1,2,3,\ldots,M\}$, $c=\{1,2,3,\ldots,N\}$ and $d=\{1,2,3\}.$
Let this be known as $CSI_{r,c,d}^{1^{\prime}}$. Each of the elements of
$CSI^{1^{\prime}}$ form the indices for position scrambling of $P$%
.\newline\newline Step B:- \newline\newline Shuffle all the $M$ columns and
$N$ rows of the original image $P(p^{R},p^{G},p^{B})$ with $CSI^{1^{\prime}}$
into the intermediate transformation matrix $T_{r,c,d}$  by the
following simple process\newline
Permutation along the rows- Here we extract the pixel value of along its rows indicated by the symbol: to form the keys for row shuffling
\[
k^{1}_{i}\leftarrow CSI^{1^{\prime}}_{:,c,d},%
\]
\[
T_{r,c,d}\leftarrow P_{k^{1}_{i},c,d}.%
\]

Next, the columns of the intermediate resultant image is also replaced by
$CSI^{1^{\prime}}$. Permutation along the column is shown by a similar procedure where the pixel values along the column are extracted to form the keys
 as under
\newline%
\[
k_{i+1}^{1}\leftarrow CSI^{1^{\prime}}_{r,:,d},%
\]%
\[
T_{r,c,d}\leftarrow P_{r,k_{i+1}^{1},d}.%
\]%

\textit{Statistical Alternation:- Confusion phase}\newline\newline Step C:-
\newline\newline In this additional phase, the matrix obtained in Eq. (13) is
used to change the binary information of the image by the XOR operation represented by the symbol $\bigoplus$. Thus,
the following step is executed to get the cipher image $CI$\newline%
\[
CI\leftarrow T\bigoplus CSI^{3}.\eqno {(14)}
\]
The operations discussed here occur along the three-color planes producing the
final encrypted image $CI$. This is the cipher image which is transmitted
along with the initial conditions as the secret key.\newline\newline%
\textbf{Decryption:-}\newline\newline Step D:-\newline\newline The same data
set $Y^{1}$ and $Y^{2}$ at the receiver end is preprocessed as in
preprocessing steps 1 to 3. Then we perform anti-XOR and reverse permutation
operation as in step A to C to recover the original image $P$. These
procedures are lossless and reversible in nature and thus, without any
overhead, we can correctly decipher the image.\newline

\section{Security analysis and simulation results}

This section uses the algorithm for encrypting a $512\times512$
colored image and examines its security. \begin{figure}[ptb]
\includegraphics[width=6.8in, height=4.0in,trim=0.0in 3in 0in 3in]{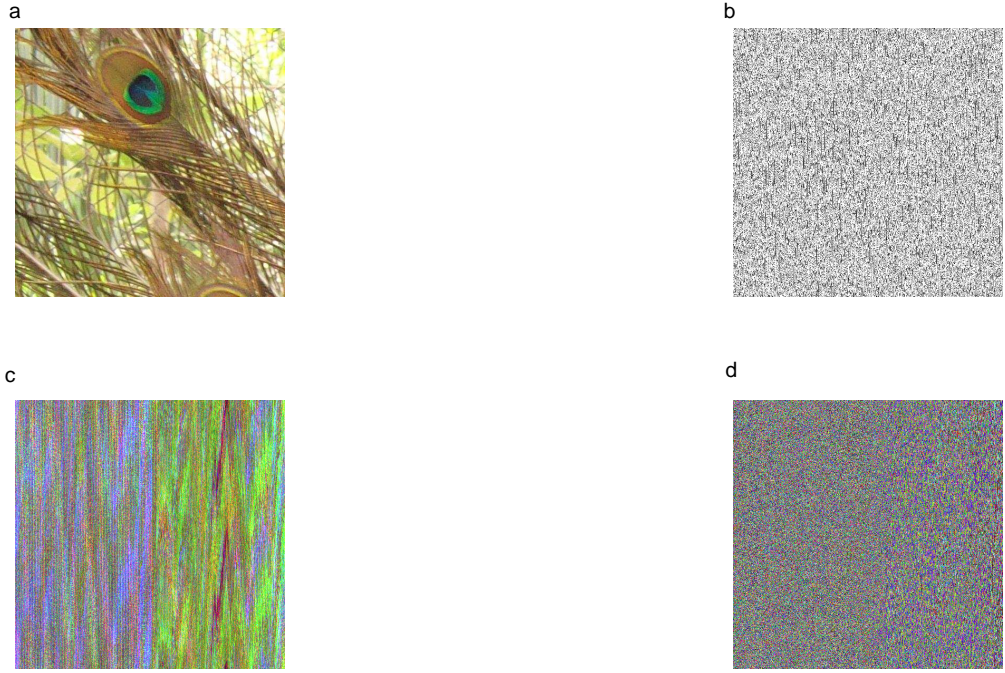}\caption{(color online) The
encryption process: (a) The original color image (b)The mask for permutation
of original image (c) The shuffled intermediate image (d) The final cipher
image. }%
\end{figure}The bands in Fig. 7 (c) results from the abrupt changes in the
diffusion process due to the inherent dynamics of the
nonlinear system under consideration. It is noteworthy, that a row or a column may be
permuted more than once which may be allowed since the computational cost in
linear check of a duplicate value in the data set is higher than the ultimate
effect of eliminating redundant position values. Figure 7(d) depicts the complete randomized
orientation of the voxels making this phase totally incoherent
from the original image.\newline

\subsection{\textit{Effectiveness of the scheme}}

The proposed method correctly recovers the original message. This is shown by
a metric known as Mean Square Error (MSE) given by
\[
MSE= \frac{1} {M \times N}\sum_{r=1}^{M}\sum_{c=1}^{N}\left[  {P(r,c)-CI(r,c)}\right]
^{2} \eqno {(15)}
\]

A correct recovery of the image will yield $MSE=[MSE^{R}\hskip2ptMSE^{G}%
\hskip2ptMSE^{B}]=[0\hskip2pt0\hskip2pt0]$ for all the three-color channels
separately. It is observed that the recovered image
highly differs from the original with a minor change in the key set as explained subsequently.\newline

\subsection{\textit{Key space attack and sensitivity analysis}}
For the numerical simulation of the system (3) and (4), we choose the initial conditions as
\[
E_{+1}(z,0)=E_{+10}+E_{+11}\cos (kz)/L,E_{+2}(z,0)=E_{+20}+E_{+22}\cos
(kz)/L ,\eqno {(16)}
\]

\[
E_{-1}(z,0)=E_{-10}+E_{-11}\cos (kz)/L,E_{-2}(z,0)=E_{-20}+E_{-22}\cos
(kz)/L, \eqno {(17)}
\]

\[
N_1=-N_{1}\cos (kz)/L, N_2=-N_{2}\cos (kz)/L. \eqno {(18)}
\]%
where \

$E_{+10}=N_{1}=0.2$, $E_{+11}=1;E_{-10}=0.3,E_{-11}=1.1;E_{+20}=0.5$, $%
E_{+22}=1.2;E_{-20}=0.6,E_{-22}=1.3, N_2=0.5, L=1/500,\epsilon=2.5 $ (in arbitrary units).\\
\\
The total number of different keys that can be utilized in a process is termed
as the key space. { The 128 bit paradigm is a widely used concept in the AES cryptography scheme which is commercially used nowadays. It is a block cipher mode under the symmetric cryptography scheme. This falls under the category of traditional cryptographic technique including DES,  RSA, DSA, hash functions, one time pad etc. However, our method based on a chaotic  symmetric cipher stream mode of encoding the message, is an offshoot of the traditional methods of cryptography. In order to render high security, it is imperative that the key space be vast enough to make any brute force attack ineffective. Also, to ensure a large divergence of a chaotic trajectory from the initial condition, the number of iterations should be relatively large. Both these issues have been addressed. The key space with its keys are generated after a considerable number of iterations which consists of the private and the public variables viz. $E_{\pm1}, E_{\pm2}$ and $N_1, N_2$.  So, due to these keys the system has a pretty exhaustive key space which is large enough and as such makes our system resistant to
brute force attacks.}
Each key has a value which lies within $0-512$ for a colored image of size $512\times512$. These are used to encode the color values of the 9 bit RGB image after tiling operation along the 3 color planes according to Eqs. (8) and (9).
\par An encryption technique is good and efficient if the
cipher image possesses sensitivity to the changes in the cipher keys, together
with a combination of sufficient large key space, in order to render the system
immune to major brute force attacks.\newline

A wrong data set in any one of the stages will result  an incorrect decrypted image as shown in Figs. 8 and  9.
The sensitivity to keys is tested on an RGB image of size $512\times
512\times3$ through the following test cases:-\newline

\begin{enumerate}
\item Test1:- Let us examine the effect of alternation of one of the secret
keys at the receiver section.The original set of keys derived from variables
used by the transmitter are $E_{\pm1}$ and $N_{1}$ from Eq. (2).  $E_{\pm1}$  =0.2 is one of the initial condition of the system which is also of the the public key. { However, the initial condition values are never transmitted over the network. Now, let us assume that an attacker intercepts the communication network and self-imposes itself as the transmitter. Since, the initial conditions are never transmitted over the network, the invader attempts to make a guess about the initial conditions. In this process, let the guessed value be assumed to be $E_{\pm1}=0.200001$.

But, the key set used for decoding at the receiver section will not match with the ones used in encoding the message as the attacker has its own wrongly generated keys. So, the effect of this tampering is reflected in the incorrect recovery of the message at the receiver section} in Fig. 8 (f).
\item Test2:- The keys in the permutation matrix Eq.  (6)  and diffusion
matrix Eq. (7) are altered through the following formulae
\[
k_{i}^{1}\leftarrow\text{integer}(\text{abs}(10^{6}\times|y_{i}^{1}%
|))\text{Mod}(M+1),\eqno{(19)}
\]

\[
k_{i}^{2}\leftarrow\text{integer}(\text{abs}(10^{2}\times|y_{i}^{2}%
|))\text{Mod}(M+1).\eqno{(20)}
\]
{ The key space increases in dimension by further increasing the dimensionality during the processing stage in eq 6. In this part, we  show that the keys so obtained are highly sensitive to slight change. Hence, if somehow or in a remotely possible case, another permutation matrix is generated by a third party who has gained an unauthorized access, the result will be an incorrect recovery of the message. So, the position permutation matrix is changed in Eq. (19) to display this effect of tampering without any assumption of changing or reducing the power of 10. Thus, the first key changes from 190 to 33, second key changes from 294 to 322, third from 398 to 362 and so on. The outcome is demonstrated in Fig 9 which also has high MSE values. Such high MSE values show that the image obtained by the receiver is not even close to the original and is highly uncorrelated from the original from which no visual interpretation can be deduced. Therefore, the receiver will come to know that the key set has been tampered. By  Eq. (20) we have shown that the keys used in diffusion phase are equally sensitive to minute disturbances on being tampered by an attacker. So, without any specific reason for the choice of the power of 2 which is raised over 10, we have modified the key space with a reduced dimension of 10 in eq 20. The outcome is shown through the MSE values listed in Table II.
Needless to say, through these exhaustive statistical tests, the sensitivity of the key sets to perturbations is validated.} The effect of these are demonstrated in Fig. 9f where the decrypted image is totally incoherent from the original and there is no way to predict any sort of information about the original message from the incorrect recovered result.The preprocessing of the key sets along with the transpose operation of the second phase mask which is then used to to form the mask image for the confusion phase adds enough stochastic properties to render the cipher image totally obscure from the original image.
\end{enumerate}

\begin{table}[th]
\caption{\textbf{Mean-square error for Test 1 and Test 2}}
\centering
\begin{tabular}
[c]{|l||lll|}\hline\hline
MSE & R & G & B\\\hline\hline
Test1 & 1.0e+003 $\times$ 9.0296 & 1.0e+003$\times$ 8.6193 & 1.0e+003$\times
$8.6450\\
Test2 & 1.0e+003 $\times$9.1335 & 1.0e+003 $\times$8.7111 & 1.0e+003 $\times
$8.5998\\[1ex]\hline
\end{tabular}
\end{table}The MSE value listed in Table II shows that incorrect decryption results in a very high MSE between the original and decrypted image. The values are not even closer to zero since the recovered output is remotely unrelated to the transmitted original image. Moreover, from Figs. 8 and 9 we can infer that it is almost impossible to
reconstruct the original image when the secret keys and the data sets are
tampered. This further reinstates that
the process is immensely sensitive to small variations to the initial conditions.
 \begin{figure}[ptb]
\includegraphics[width=6.5in, height=3.5in,trim=0.0in 2in 0in 3in]{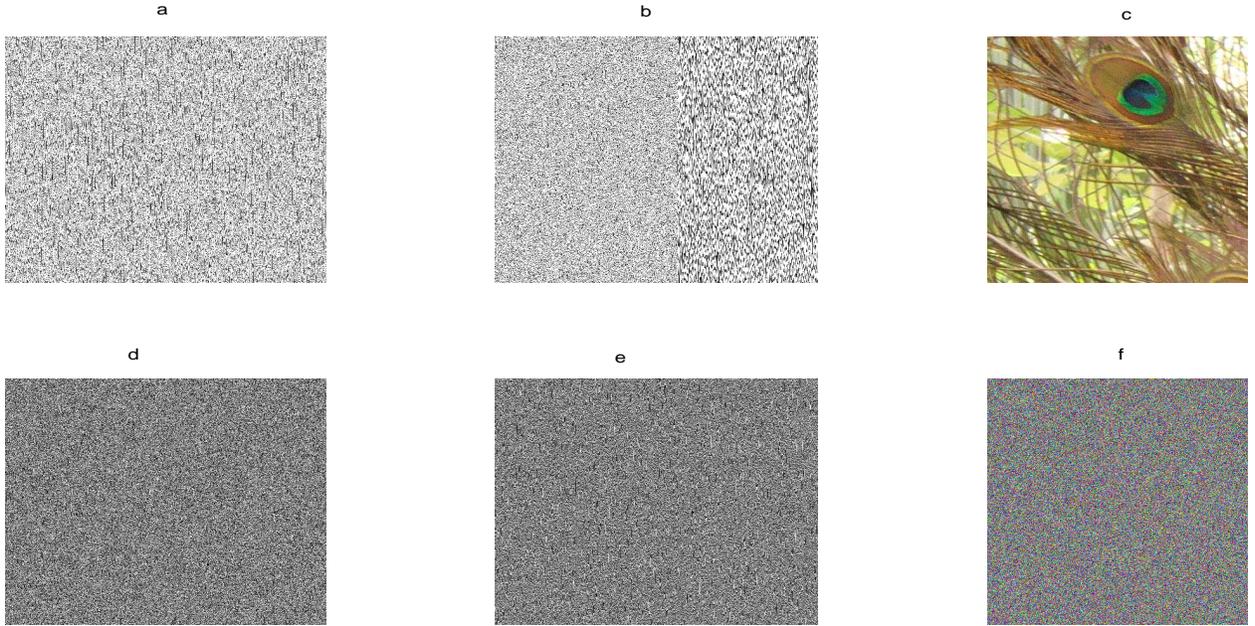}.\caption{(color online) Test
result of statistical Test 1: (a) Chaotic phase mask $CSI^{1}$ for the
permutation phase with the correct keys and initial conditions (b) Second
chaotic phase mask $CSI^{2}$ obtained from the correct keys and initial
conditions (c) Correct decrypted image (d) First chaotic phase mask for the
permutation stage obtained from the incorrect keys and tampered initial,trim=0.0in 0in 0in -1.5in
conditions (e) Second chaotic phase mask obtained from the incorrect keys and
tampered initial conditions (f) Incorrectly decrypted image }%
\end{figure}

\begin{figure}[ptb]
\includegraphics[width=6.5in, height=3.5in,trim=0.0in 2in 0in 3in]{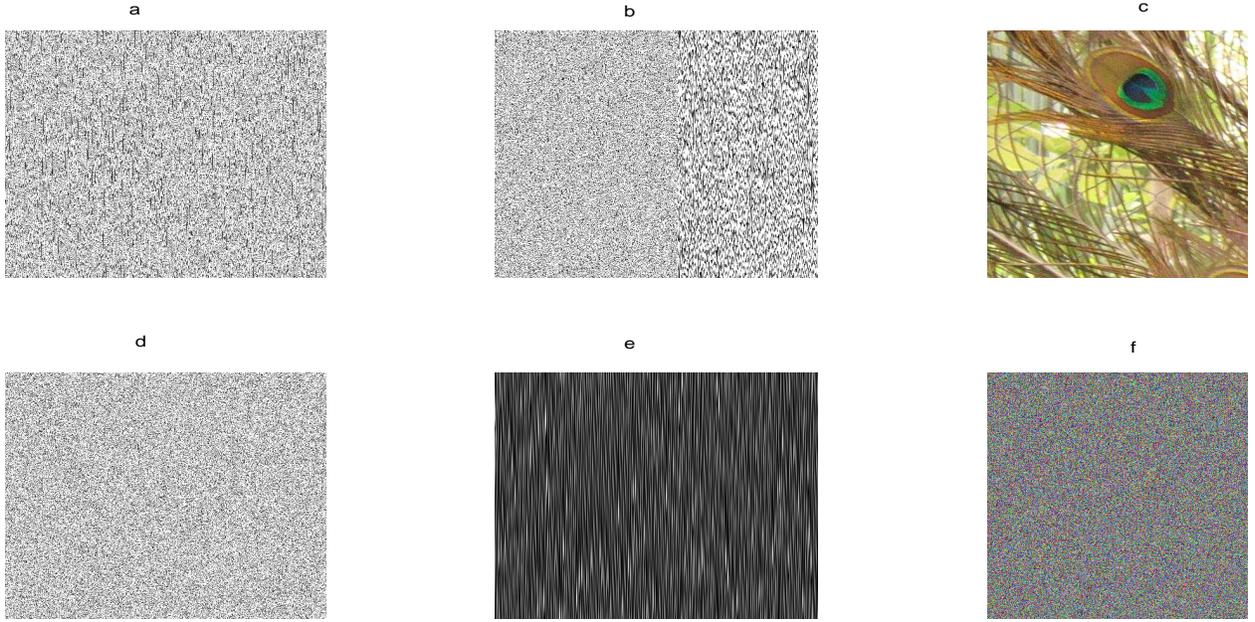}.\caption{(color online) Test
result of statistical Test 2: (a) Chaotic phase mask $CSI^{1}$ for the
permutation phase with the correct keys (b) Second chaotic phase mask
$CSI^{2}$ obtained from the correct keys (c) Correct decrypted image (d) First
chaotic phase mask for the permutation stage obtained from the incorrect keys
(e) Second chaotic phase mask obtained from the incorrect keys (f) Incorrectly
decrypted image. }%
\end{figure}

\subsection{\textit{Differential attack resistance}}

A good cipher should be immune to the influence of a change in a pixel.
The rate of change of the number of pixels of the cipher image $CI$ when the plain
image $P$ is altered by just one pixel is quantified by a metric known as NPCR.
It is defined by
\newline%

\[
NPCR=\frac{\sum_{r=1,c=1}^{M,N}D(r,c)}{M\times N}\times100,\eqno {(21)}
\]
where $D(r,c)$ represents the change in the picture element due to encryption for a monochrome image.
$$ D(r,c)=0 \hskip 2 pt when \hskip 2 pt P(r,c)=CI(r,c)else $$
$$ D(r,c)=1 \hskip 2 pt when \hskip 2 pt P(r,c)\neq C(r,c)$$
In our example, the change in a few voxel values before and after encryption are listed below to demonstrate the dispersion of
voxels through the scheme\newline%
\begin{align}
P(1,1,3)  & =153,CI(1,1,3)=218,\nonumber\\
P(2,2,3)  & =148,CI(2,2,3)=74,\nonumber\\
P(50,50,3)  & =152,CI(50,50,3)=113,\nonumber\\
P(100,100,3)  & =79,CI(100,100,3)=238,\nonumber\\
P(256,256,3)  & =42,CI(256,256,3)=123.\nonumber
\end{align}
This shows that the resulting cipher image has its picture elements totally changed. Next quantifiable measure of diffusion
properties is the Unified Average Changing Intensity (UACI)which determines the
average intensity of the differences between the pixel values of the original
and encrypted image. It is expressed by\newline%
\[
UACI=\frac{1}{M\times N}\sum_{r=1,c=1}^{M,N}\frac{|P(r,c)-CI(r,c)|}{255}%
\times100. \eqno {(22)}
\]
The results of NPCR and UACI in percentage is shown in Table III. From it we can deduce that the scheme has a high value of NPCR along with a satisfactory value of UACI. Thus, higher the value of NPCR, more is the system resistant to
differential attacks.
This validates the efficiency of the proposed scheme.
\begin{table}[th]
\caption{Result of differential attack: NPCR and UACI performance metric}
\centering
\begin{tabular}
[c]{|l||lll|}\hline\hline
& R & G & B\\\hline\hline
NPCR & 99.61 & 99.62 & 99.60\\
UACI & 14.47 & 6.97 & 10.35\\[1ex]\hline
\end{tabular}
\end{table}
\subsection{\textit{Histogram analysis}}

The distribution of the pixels after being scrambled or manipulated can be
studied with the help of a chart which represents the distribution of the
pixels in the range $0-255$. For a gray image having gray level $\mathit{r_{q}%
}$ it is represented by a function hist$(l_{q})=n_{q}$, where $l_{q}$ denotes
the $\mathit{q}^{th}$ gray level and $\mathit{n_{q}}$ are the number of pixels
in an image. In the diffusion phase, the positions of the image elements
undergo shuffling. This does not alter the statistical information of the
original image. However, the additional layer of security will disguise the
desired information, as reflected in the histogram of Fig. 10. 
\begin{figure}[ptb]
\includegraphics[width=6.5in, height=3.5in,trim=0.0in 2in 0in 3in]{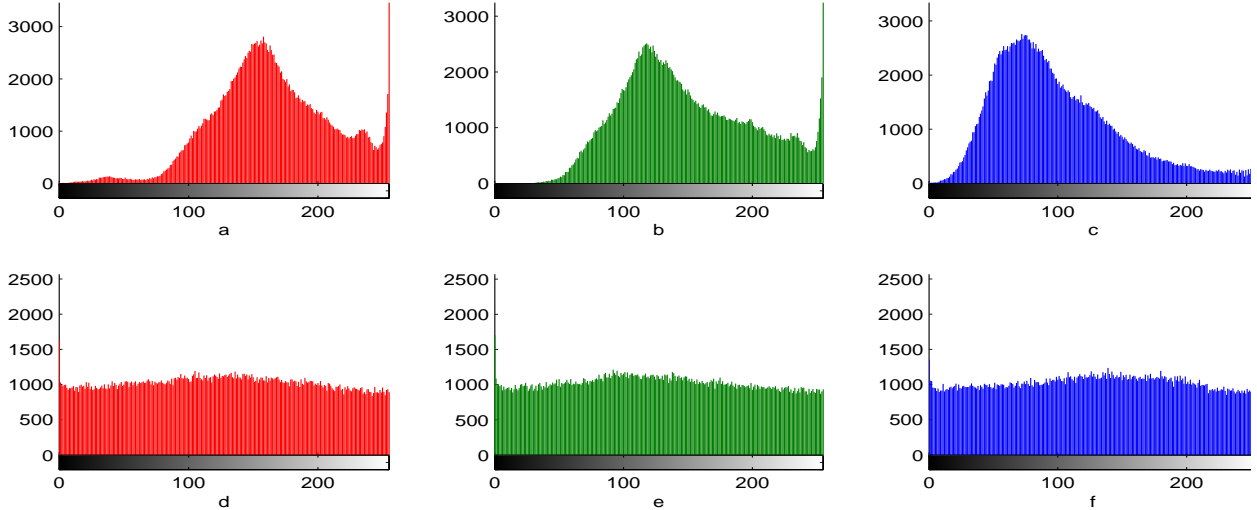}\caption{(color online) Histogram:
(a)-(c) original image, (d)-(e) cipher image. }%
\end{figure}The intensity of the encrypted image greatly differs from the
original image, thus an attacker will be unable
to infer any statistical information required to decode from the scheme. From the result of intensity distribution, it can be ascertained that the
scheme also possesses good confusion properties.

\subsection{\textit{Correlation analysis of pairs of adjacent pixels}}

Statistical analysis of the cipher image can be performed by another excellent
metric, the correlation coefficient. This shows whether there is any occurrence of an
association between the guessed secret key and the observed information of the
cryptosystem and forms the core information for cryptanalysis. The
covariance cov$(x,y)$ between a pair of pixel values present in the subset $x$ and $y$ in a gray
image is defined by\newline%
\[
\text{cov}(x,y)=E(x-E(x))(y-E(y)),
\]
Then, the correlation coefficient $r_{xy}$ is given by\newline%
\[
r_{xy}=\frac{Cov(x,y)}{\sqrt{D(x)}\sqrt{D(y)}}. \eqno {(23)}
\]

where $E(x), E(y)$ denotes the mean; $D(x),D(y)$ stands for the variance between the pixels.
Correlation coefficient values of a random sample of $1000$ voxels, placed
adjacently along the horizontal, vertical and diagonal directions, have been
computed for each of the three-color channels and enumerated in Table IV. From
the values, it can be observed that the correlation of cipher image is almost zero.
For an efficient encryption methodology, it is imperative that the correlation
values of adjacent pixels be minimal for the encoded image. Our scheme
fulfills this criteria. 
\begin{table}[th]
\caption{Correlation Coefficient for each color band}%
\label{Table:Correlation Coefficient}
\centering
\begin{tabular}
[c]{|l||l|l||l|l||l||l|}\hline
& \multicolumn{3}{l|}{Original} & \multicolumn{3}{l|}{Cipher}\\\cline{2-7}
& R & G & B & R & G & B\\\hline\hline
Horizontal & 0.9714 & 0.9756 & 0.9730 & -0.0002 & -0.120 & 0.0207\\
Vertical & 0.9592 & 0.9661 & 0.9610 & 0.0429 & 0.0539 & 0.0793\\
Diagonal & 0.9621 & 0.9690 & 0.9640 & 0.0305 & 0.0270 & 0.0281\\\hline
\end{tabular}
\end{table}\begin{figure}[ptb]
\includegraphics[width=6.5in, height=3.5in,trim=0.0in 2in 0in 3in]{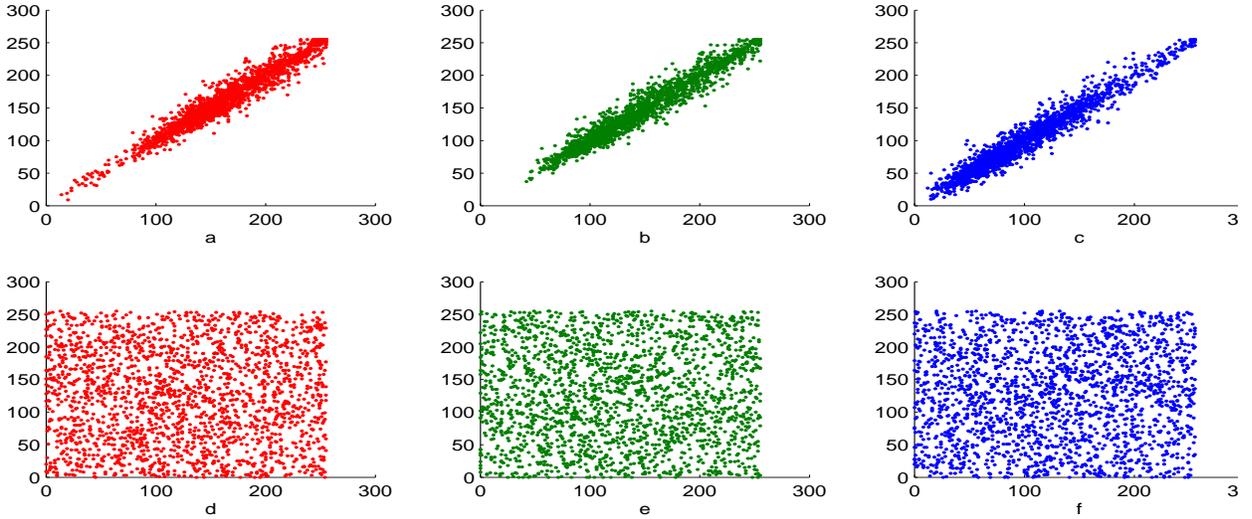}\caption{(color online) Scatter
plot of correlation coefficients of adjacent pairs of pixels along the
horizontal direction: (a) Correlation coefficient of the original image in the
Red color plane (b) Correlation coefficient of the original image in the green
color plane (c) Correlation coefficient of the original image in the blue
color plane.}%
\end{figure}Fig. 11 displays the spatial distribution of the randomly selected
voxels from the three-color planes placed adjacently along the horizontal
direction. The graphical result emphasizes that there is hardly any
correlation between the original and the distorted version of the image.

\section{Conclusion}

The semiconductor laser represents a good amplifier configuration which exhibits stable as well as strong chaotic turbulent states. The
synchronization of semiconductor lasers in their chaotic regime is still a
open area of research. The corresponding  system of PDEs
is, infinite dimensional  and continuous coupling terms necessary for the synchronization. The coupling directly injects the output power from driving system to
response system. Numerical simulations show that the complete synchronization
is achieved although both the systems are chaotic in nature.\newline We
have used the dynamics of the synchronized states of the system, which
play the roles of the transmitter and receiver sections, in colored image
cryptography. The image can be securely
transmitted and recovered over the secure communication channel. The process is applied to a colored image affecting the three color planes. The key space
is vast enough that attackers may very hardly guess a key. Also, the speed of
encryption is notable since the key set preprocessed
before hand. The second phase of encryption
with a chaotic phase mask renders the system immune to common
statistical attack, even if the initial conditions are identified thereby,forming a protective shield. Furthermore, the scheme can be extended by encrypting several
image maps of the chaotic data set, thereby adding several layers of security.
The proposed scheme has been tested exhaustively. Statistical tests re-instate that the scheme is also sensitive to its keys and it is almost impossible to infer any information about the original image if the chaotic set is tampered with. Overall, our method is robust and found to be resistant to
most of the common threats.
\acknowledgments{LR acknowledges partial support form the European Research Council, under the
European Community's Seventh Framework Programme (FP7/2007-2013) / ERC grant
agreement n 202680. The EC is not liable for any use that can be made on the
information contained herein. APM acknowledges support from the Kempe Foundations, Sweden.}

\end{document}